          [1999/12/02 1.01 (PWD)]
\documentclass[a4paper,twocolumn]{esapub} % European paper

\usepackage{epsfig}
\usepackage{natbib}
\title{The Cygnus Region: $^{26}{\rm Al}$ from OB Associations}
\author{S. Pl\"uschke}
\author{K. Kretschmer}
\author{R. Diehl}
\affil{Max-Planck-Institut f\"ur extraterrestrische Physik, 85740 Garching,
Germany}
\author{D.H. Hartmann}
\affil{Dep. of Physics \& Astronomy, Clemson University, Clemson, SC 29634, USA}
\author{U.G. Oberlack}
\affil{Astrophysics Laboratories, Columbia University, New York, NY 10027, USA}
\newcommand{\Al}{\ensuremath{^{26}{\rm Al}} }

\newcommand{\Fe}{\ensuremath{^{60}{\rm Fe}} }

\newcommand{\Msun}{\ensuremath{{\rm M}_{\odot}} }

\begin{document}


\maketitle

\begin{abstract}
The COMPTEL map of the 1.809 MeV gamma-ray line, which is attributed
to the radioactive decay of $^{26}$Al, shows significant excess emission
in the Cygnus region. Cygnus is a region of recent star formation activity,
which is rich in massive, early-type stars concentrated in OB associations.
We perform population synthesis studies of this region to model the production 
and distribution of $^{26}$Al, which is predominantly synthesized in massive
stars and distributed through stellar winds and core-collapse supernovae.\\
Our simulations of the Cygnus OB associations determine the time-dependent
production rate of $^{26}$Al. We also estimate the associated production of
radioactive $^{60}$Fe, which is primarily synthesized in core-collapse
supernovae. Gamma-ray line emission from this isotope has yet to be detected.
We also track the temporal evolution of the mechanical "luminosity" due to the
kinetic energy of stellar winds and supernova ejecta, and the ionizing EUV 
emission from early-type stars. Together with a simple one-dimensional model 
of the superbubble driven into the ISM by this energy flux one can constrain 
key parameters of the stellar associations in the Cygnus region.
\end{abstract}

\section{The Cygnus region}
\begin{figure}[h!]
  \begin{center}
    \epsfig{figure=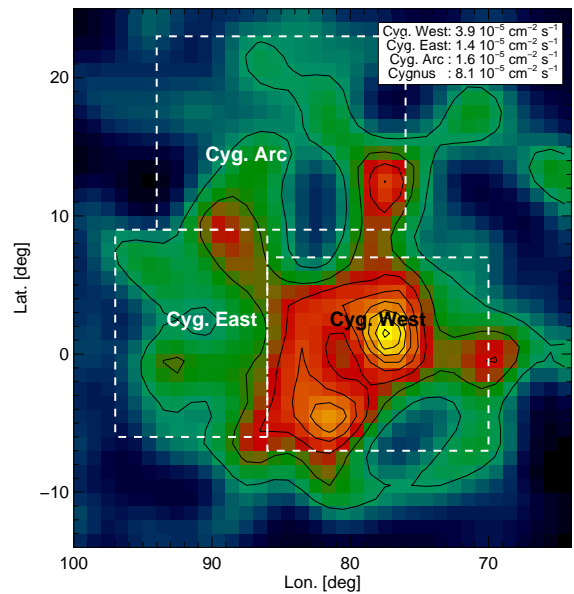,width=8cm}
    \caption{Close-up of the Cygnus region from the most recent COMPTEL 1.809 MeV 
     maximum entropy map. The deduced fluxes for different areas are given in the insert.}
    \label{fig:me}
  \end{center}
\end{figure}

Recent COMPTEL maps of the galactic 1.809 MeV emission show enhanced emission
in the Cygnus region. Figure \ref{fig:me} displays a detailed image of this
region on the sky derived with the maximum entropy imaging method \citep{plueschke00a}. 
The flux peaks strongly in the area labelled {\em Cygnus West} and extends over 
$\sim 25^{\circ}$.\\
Extensive comparisons of the 1.809 MeV map to potential tracers \citep{knoedl99} strongly 
support the idea that massive stars (e.g. \cite{meynet97}) and their sub-sequent core-collapse
supernovae (e.g. \cite{ww95}) are the dominant sources for interstellar \Al. In addition,
the studies of \cite{kolb00}, based on detailed modeling of binary star populations, suggest
that the integrated contribution from novae to the galactic \Al budget is most likely less
than $0.1\,\Msun$, i.e., less than 10$\%$ of the total amount inferred from the 1.809 MeV map.
A comparison with other determinations \citep{s93} renders the estimation of the nova contribution
to be seriously affected by theoretical uncertainties. Nevertheless, due to the much longer
stellar evolution the nova contribution could be neglected in our attempt to model OB
associations. The contribution of \Al from AGB stars \citep{mm00} is expected to be less than
$\sim 0.4\,\Msun$. It is presently impossible to distinguish between contributions from
hydrostatic burning, injected by stellar winds, and \Al from explosive burning taking 
place in the shocks of core-collapse supernovae. Only the strong correlation of the observed
1.809 MeV emission with tracers of massive stars (most significant for free-free emission)
provides some clues on the relative importance of the two production channels.\\
The Cygnus region contains 23 known Wolf-Rayet stars \citep{vdh00}, of which
14 are classified as WN stars, 8 as WC and 1 as WO type. In addition, recent
supernova remnant (SNR) catalogues list 19 SNR in this area \citep{snr00}.
Furthermore, 7 OB associations have thus far been associated with the Cygnus region.\\
We have extended traditional stellar population synthesis models by including the 
nucleosynthesis of \Al and \Fe in massive stars and supernova explosions. This allows
us to determine the amount of radioactive material as a function of time for a given
star formation history. In addition to calculating the budget of radioactivity in the
star forming region, population synthesis models also provide the energy production rate
due to the kinetic energy in stellar winds and expanding supernova ejecta, as well as
the flux of ionising radiation from hot, massive stars. It is thus possible, in principle,
to reconstruct, or at least constrain, the star formation history of the Cygnus region
by combining data on the dynamics, radiation environment, and gamma-ray line fluxes.

\section{The OB association model}
\begin{figure}[h!]
  \begin{center}
    \epsfig{figure=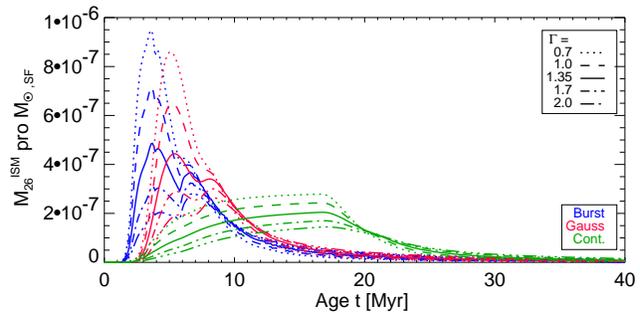,width=8.0cm}
    \caption{Evolution in time of the amount of \Al for various assumptions
      about the IMF slope and the SFR(t) profile.}
    \label{fig:m26}
  \end{center}
\end{figure}

\begin{figure}[h!]
  \begin{center}
    \epsfig{figure=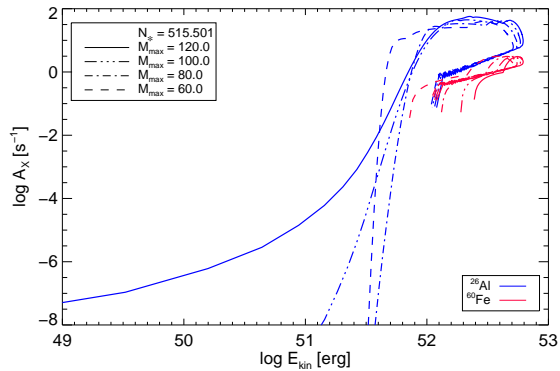, width=8cm}
    \caption{Activity (decay rate) of \Al \& \Fe in a star forming region and the 
        associated kinetic energy of the expanding supershell, driven mostly by the stars 
        that are also the producers of these radioactive isotopes.} 
    \label{fig:lwact}
  \end{center}
\end{figure}

Our population synthesis model is based on the idea of continuous parameter
distributions \citep{plueschke99} and was applied to the Cygnus region
by \cite{plueschke00b}. Based on the assumption of a time-invariant initial
mass function (IMF) and a mass-invariant star formation rate function,
SFR(t), we compute models for OB associations of different richness,
duration of star forming activity, and slope of the initial mass function. 
In all cases the IMF is assumed to be a single power-law with mass limits
$M_{low}=8\,\Msun$ and $M_{up}=120\,\Msun$, and the parameter $\Gamma$
determines the slope via $\Phi\propto M_{\rm ZAMS}^{-(1+\Gamma)}$.\\ \\
For the computation of the time-dependent ejection of \Al and \Fe due
to stellar winds and core-collapse supernovae we use the predicted yields
from the detailed nucleosynthesis models of \cite{meynet97} and \cite{ww95}.
For a given star formation history the production rate of \Al and \Fe follows
from stellar life times and the time- and mass-dependent yields. Radioactive decay
then predicts the emerging gamma-ray line flux together with the accumulated mass
of \Al and \Fe, which is displayed as a function of time in figure
\ref{fig:m26} for the case of \Al.
The displayed profiles are normalised to the total mass converted into stars in the
given mass range, varying the slope of the IMF and SFR(t). As long as the IMF is not
too flat, the time profiles of populations with short duration star formation are
characterised by two peaks. The first peak is attributed to the stellar wind
component, whereas the second peak is due to the delayed supernova contribution.
For very flat IMFs the profiles are very strongly peaked in the early phase.
Long duration episodes of star formation (keeping the total amount of stars produced
constant) result in flat profiles (essentially a situation of steady state) with
significantly reduced values for the maximum amount of radioactivity.\\ \\
Through their enormous mass loss rates, in particular during the Wolf-Rayet phase,
massive stars release a large amount of kinetic energy into the surrounding medium.
Sub-sequent supernovae contribute a large amount of kinetic energy through the 
ejecta moving with initial speeds in excess of 10000 km/s. By applying the Geneva
mass loss description \citep{meynet97}, a semi-empirical velocity law
\citep{how89,pri90}, and assuming the canonical $10^{51}$ erg per supernova explosion,
we calculated the evolution of the mechanical luminosity of an OB association.
The kinetic energy ejected by an association can in principle be observed in
form of an expanding supershell around the association. Several processes
coupled to the shell expansion modify the observable kinetic energy residing 
in these shells. To link the observed kinematic properties of these shells to the
underlying energy input rate we extended our population synthesis model
by an one-dimensional bubble expansion model. This model uses the thin shell
approximation \citep{ss95} and includes radiative cooling and mass loading of 
the bubble interior due to partial shell evaporation. We find that the kinetic 
energy derived from observations of the shell is typically less than 20$\%$
of the injected energy. However, this fraction depends strongly on the details of the
poorly understood evaporation process taking place at the interfaces between shell
material and the hot bubble interior. More efficient mass loading of the bubble 
interior increases the density, which leads to more efficient radiative cooling.
Efficient shell evaporation thus causes the expansion to stall sooner.
Figure \ref{fig:lwact} displays the evolution of the relation between the activity
(decay rate) of the radio isotopes and the kinetic energy of the shell. Both quantities 
can be obtained observationally, thus providing a strong constraint SFR(t).
 \\ \\
%\newpage
Fits to tracer maps \citep{knoedl99} have shown that free-free emission from the
ionised component of the ISM is most strongly correlated with the 1.809 MeV map.
This motivates the inclusion of the emission of photoionizing EUV radiation from
massive, young stars in our population synthesis model. Based on the
calculations of \cite{vacca96} we used a simple fit function to compute
the integrated emission rate. Our simplified description overestimates
the EUV emission by up to 0.3 dex because the effects of stellar evolution are
not included rigorously. One of the main features in the evolution of the ionizing
flux is its rapid decrease, due to the fact
that only the most massive stars contribute significantly to this
radiation. Observations during the early phase of a starburst thus provide  
valuable constraints on the stellar population near the upper end of the IMF. This 
is also the regime probed by gamma-ray line observations. 
\section{Cygnus OB associations}
Based on the OB associations listed in \cite{boch85} we applied our 
OB association model to the gamma-ray observations of the Cygnus region.
We attempt to constrain model parameters for this region by directly using 
star count data, as well as age estimates and shell energy determinations
from the literature. For the $\gamma$-ray intensity model of the Cygnus region we
assumed a circular shape of the emission areas with a central plateau ranging over
2/3 of the angular radii of the associations and a sharp Gaussian drop-off near the
border.
\begin{figure}[h!]
  \begin{center}
    \epsfig{figure=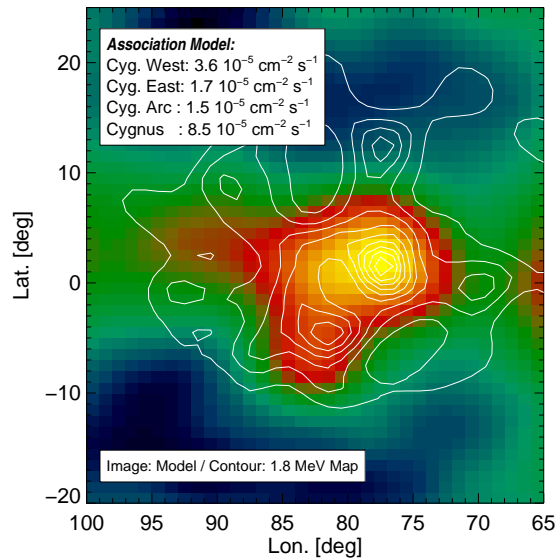,width=8cm}
    \caption{Comparison of the reconstructed maximum entropy image of
             the association (image) with the COMPTEL 1.8 MeV
             map (contour).}
    \label{fig:mod}
  \end{center}
\end{figure}
Fitting of this OB association model with six independent associations to the
COMPTEL 1.8 MeV data attributes the strongest contribution to Cygnus OB2
followed by a 50\% smaller contribution from Cygnus OB1. In this model the
remaining OB associations contribute less than 25\% to the total 
flux. Figure \ref{fig:mod} shows an overlay of the maximum entropy
image (contours; see Fig. \ref{fig:me}) and a maximum entropy deconvolution
reconstruction of the fitted model (image).\\
This result is consistent with the recent work of \cite{knoedl00}, who
found that Cygnus OB2 is the richest OB association known to date.
By using the 2MASS survey data \cite{knoedl00} concluded that Cygnus OB2
resembles more closely to a young globular cluster, such as the Arches cluster,
than a typical OB associations. He found about 120 O star and 2600 OB star
members in an area of $2^{\circ}$ in diameter. The large number of OB
stars revives the idea \citep{ct94} that Cygnus OB2 triggered the star
formation activity in Cygnus OB1, OB3, OB7 and OB9.
\cite{c98} concluded from kinematic arguments that Cygnus OB2 must provide a
steady-state mechanical luminosity of $\sim4.7\cdot10^{39}\,{\rm erg\,s^{-1}}$,
to get this scenario working. This implies that 2800 early type stars should 
reside in Cygnus OB2, which is in agreement with the observation of \cite{knoedl00}.\\
However, if one regards results from an the analysis of a continuum radio survey of
the Cygnus X region by \cite{wendker91}, which results in a none detection of new
supernova remnant inside Cygnus OB2, a contradiction arises. G78.2+2.1 appears to
be the only SNR in this field, which in turn implies that Cygnus OB2 is significantly
younger than 4.5 Myr. This conclusion is supported by the large number O stars found
by \cite{knoedl00} and by the HR-diagram extracted by \cite{massey95}, which gives
an deduced age of $2.5\pm1.0\,{\rm Myr}$ for Cygnus OB2. The absence of supernova
activity inside Cygnus OB2 forces the number of early stars to be four times higher
to fulfill the energy criterion for propagating star formation, which is in contradiction
to the observations of \cite{knoedl00}. The implications of our model fit (that
Cygnus OB1 may provide a larger amount of \Al) supports this argument due to the
fact that OB1 must be old enough to release \Al, which is not the case in the
propagation scenario assuming an age of less than 4~Myr for Cygnus OB2.
Therefore we are inclined to disregard the idea of propagating star
formation by an expanding bubble structure around Cygnus OB2.\\
However, supernovae appearing inside hot, thin cavities may not
produce visible remnants so the constrain on OB2's age may not be
very stringent. If indeed OB2 is older than previously assumed the
proposed scenario may work.
\section{Uncertainties}
Our model is subject to two different types of uncertainties. First of
all, all of the input components of the model have systematic uncertainties 
in the theoretical descriptions used to build the basis of the model.
For example, in the case of the nucleosynthesis part, the employed stellar
evolution models are still hampered by an incomplete understanding of 
internal mixing processes in stars. The description of convection has
been intensively debated over recent years. In addition, uncertainties
in the mass loss description as well as in the rotation treatment strongly
affect the yield estimates. Recent investigations of stellar evolution
\citep{meynet00} and sub-sequent core-collapse supernovae from rotating
stars \citep{woosley99} have shown stellar rotation to be a key ingredient in
understanding the coupling of internal mixing and mass loss. One of the most
severe changes due to rotation is a extension of the main sequence phase of up to
25\% compared to non-rotating models (using same input physics). However, \cite{meynet00}
conclude that their earlier models \citep{meynet97} using Schwarzschild convection
with moderate overshooting together with an enhanced mass loss may reasonably describe
the behaviour of massive stars rotating with a mean velocity and avoiding overshooting
and mass loss enhancement. Nevertheless, the extended main sequence lifetime of
rotating stars may alter the shape of the predicted ISM abundances by shifting
and expanding it in time. In addition, in some cases the nuclear input to the employed
reaction networks is not sufficiently well known to derive accurate yields.
In summary, each imperfection of the input models may give rise to individual
uncertainties of up to a factor 2.\\
In the case of \Al \cite{langer98} demonstrated that \Al synthesis can be strongly
affected by the presence of a binary companion. Their simulations suggest possible 
yield enhancements by factors up to 1000. Clearly the interpretation of the
observed 1.809 MeV emission in the Cygnus region, and the Galaxy as a whole, 
depends on our understanding of this binary star "correction factor". More 
stellar evolution studies are clearly needed in this area. We assume that significant
uncertainties are induced by our lack of a full understanding of nucleosynthesis
in binary star systems, but it is presently not possible to obtain reliable estimates
of these uncertainties.\\
Because stellar mass loss is still poorly understood \citep{lc99}, predicting 
the mechanical luminosity is uncertain as well. \\
Also the determination of the initial mass function and the star formation
history is subject to uncertainties. For our standard case we are using
the Salpeter-IMF. \cite{knoedl00} determined a steeper IMF slope
of $\Gamma=1.6\pm0.1$ for Cygnus OB2, which represents a slope as favoured by
\cite{kroupa00}.\\ \\
In addition to these systematic uncertainties statistical fluctuations due
to the finite population size of real OB associations may affect the
predicted light curves. To include this aspect in our model we use a Monte Carlo 
version of our population synthesis code.
\begin{figure}[h!]
  \begin{center}
    \epsfig{figure=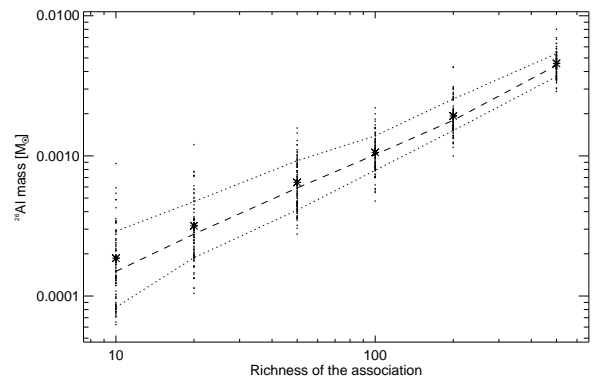,width=8cm}
    \caption{Dependence of spread of the interstellar \Al mass as function
     of the richness (number of stars more massive than $8\,{\rm M}_{\odot}$)
     of the association.}
    \label{fig:rich}
  \end{center}
\end{figure}\\
Figure \ref{fig:rich} shows the dependence of the simulated spread of the 
amount of \Al (measured at the time
of peak 1.809 MeV emission) on the richness of the association. For 
populations containing more than 100 stars the statistical uncertainties become
less than the systematical ones.\\
\section{Summary \& Prospects}
We have added the production of radioactive \Al and \Fe to a population synthesis
code in order to study gamma-ray line emission from Galactic star forming regions.
The numerical simulation of a star burst (or a more extended period of star forming
activity) allows us to calculate time-dependent gamma-ray line fluxes, which can be 
compared to observations to constrain key parameters, such as the strength, duration,
and shape of the SFR(t) function. However, the model presented here is specifically
designed for applications to OB associations, i.e., star formation involving a rather
limited number of stars. We therefore had to investigate statistical uncertainties
with Monte Carlo methods. For OB associations containing fewer than $\sim$ 100 stars
statistical uncertainties dominate.\\
Application of our model to the Cygnus region demonstrates the use of gamma-ray
observations as a complementary tool in the study of star formation activity in the
Milky Way. Subtracting a contribution of approximately 40\% of the total observed
flux due to isolated sources such as WR stars and SNRs we find Cygnus OB2 to be the
dominant source of interstellar \Al in this region contributing about 50\% of the
integral gamma-ray line emission attributed to the Cygnus associations. Cygnus OB1
contributes up to 25\% whereas the remaining associations altogether add another 25\%.
Our population synthesis disregards the propagating star formation hypothesis due to
some contradictions found in the expected gamma-ray line flux and the kinetic energy
output.\\ \\
Future studies with our population synthesis tool will utilize constraints from 
the observable kinetic energy in an expanding supershells. To take advantage of 
this particular diagnostic tool, a careful study of the effects of shell evaporation 
needs to be carried out. We plan to use existing HI surveys to obtain data for a larger
sample of Galactic supershells, to determine properties of these shells, and to progress
to population synthesis simulations of global star formation in the Galaxy. Based on the
interpretation of existing gamma-ray line observations, such as the COMPTEL \Al map at 
1.809 MeV, we will simulate the performance of the INTEGRAL mission, especially with 
focus on the gamma-ray lines from \Fe that have yet to be detected.


\begin{thebibliography}{}
\bibitem[Bochkarev \& Sitnik(1985)]{boch85}
Bochkarev N.G., Sitnik T.G., 1985, A\&SS 108: 237
\bibitem[Comer\'on \& Torra(1994)]{ct94}
Comer\'on F., Torra J., 1994, ApJ 423: 652
\bibitem[Comer\'on et~al.(1998)]{c98}
Comer\'on F., et~al., 1998, A\&A 330: 975
\bibitem[Green(2000)]{snr00}
Green, D.A., 2000, {\it Cat. of gal. SNR (August 2000)},
URL: {\it http://www.mrao.cam.ac.uk/surveyz/snrs/}
\bibitem[Howarth \& Prinja(1989)]{how89}
Howarth I.D., Prinja R.K., 1989, ApJS 69: 527
\bibitem[Kn\"odlseder et~al.(1999)]{knoedl99}
Kn\"odlseder J., et~al., 1999, A\&A 344: 68
\bibitem[Kn\"odlseder(2000)]{knoedl00}
Kn\"odlseder J., 2000, A\&A 360: 539 - 548
\bibitem[Kolb (2000)]{kolb00}
Kolb U., 2000, contr. to  the workshop {\it The Impact of Binaries on Population Studies},
Brussels, 19.-25.8.2000, in prep.
\bibitem[Kroupa et~al.(2000)]{kroupa00}
Kroupa P., et~al., 2000, astro-ph/0009005
\bibitem[Lamers \& Cassinelli(1999)]{lc99}
Lamers H.J.G.L.M., Cassinelli J.P., 1999, "Stellar Winds" Cambridge Univ. Press
\bibitem[Langer \& Braun(1998)]{langer98}
Langer N., Braun H., 1998, Proc. of the $9^{th}$ Workshop of Nuclear Astrophysics,
MPA Garching, p. 18
\bibitem[Massey et~al. (1995)]{massey95}
Massey P., et~al., 1995, ApJ 454: 151
\bibitem[Meynet et~al.(1997)]{meynet97}
Meynet G., et~al., 1997, A\&A 320: 460
\bibitem[Meynet \& Maeder(2000)]{meynet00}
Meynet G., Maeder A., 2000, A\&A 361: 101 - 120
\bibitem[Mowlavi \& Meynet(2000)]{mm00}
Mowlavi N., Meynet G., 2000, A\&A 361: 959 - 976
\bibitem[Pl\"uschke et~al.(1999)]{plueschke99}
Pl\"uschke S., et~al., 1999, in Proc. of the workshop {\it Astronomy with Radioactivities},
Tegernsee, p. 197 - 204, MPE Report 274 (ISSN 0178-0719)
\bibitem[Pl\"uschke et~al.(2000)]{plueschke00a}
Pl\"uschke S., et al., 2000a, {\it The COMPTEL 1.809 MeV Survey} in these proceedings
\bibitem[Pl\"uschke et~al.(2000)]{plueschke00b}
Pl\"uschke S., et~al., 2000b, Proc. of the $5^{\rm th}$ Compton Symposium,
AIP Conf. Proc. Vol. 510, p. 44 - 48
\bibitem[Prinja et~al.(1990)]{pri90}
Prinja R.K., et~al., 1990, ApJ 361: 607
\bibitem[Shull \& Saken(1995)]{ss95}
Shull J.M., Saken J.M., 1995, ApJ 444: 663 - 671
\bibitem[Starrfield et~al.(1993)]{s93}
Starrfield S., at~al., 1993, Phys. Rep. 227: 223 - 234
\bibitem[Vacca et~al.(1996)]{vacca96}
Vacca W., et~al., 1996, ApJ 460: 914
\bibitem[van der Hucht et~al.(2000)]{vdh00}
van der Hucht K., et~al., 2000, {\it VII. gal. WR Catalogue}, accepted for publ.
in New Astr. Rev.
\bibitem[Wendker et~al.(1991)]{wendker91}
Wendker et~al., 1991, A\&A 241: 551
\bibitem[Woosley \& Weaver(1995)]{ww95}
Woosley S.E., Weaver T.A., 1995, ApJS 101: 181
\bibitem[Woosley \& Heger(1999)]{woosley99}
Woosley S.E., Heger A., 1999, in Proc. of the workshop {\it Astronomy with Radioactivities},
Tegernsee, p. 133 - 140
\end{thebibliography}
\end{document}